\documentclass[english,aps,manuscript]{revtex4}
\usepackage[T1]{fontenc}
\usepackage[latin9]{inputenc}
\usepackage{amsmath}
\usepackage{graphicx}
\usepackage{amssymb}
\usepackage{esint}

\usepackage{babel}

\begin{document}

\title{Double humped states in the nonlinear Schr\"{o}dinger equation with
a random potential}

\author{H. Veksler, Y. Krivolapov and S. Fishman}
\begin{abstract}
The role of double humped states in spreading of wave packets for
the nonlinear Schr\"{o}dinger equation (NLSE) with a random potential
is explored and the spreading mechanism is unraveled. Comparison with
an NLSE with a double-well potential is made. There are two independent
affects of the nonlinearity on the double humped states for the NLSE:
coupling to other states and destruction. The interplay between these
effects is discussed.
\end{abstract}
\maketitle
We consider the discrete nonlinear Schrodinger equation with a random
potential in one dimension:\begin{equation}
i\frac{\partial\psi_{n}}{\partial t}=-\psi_{n+1}-\psi_{n-1}+\epsilon_{n}\psi_{n}+\beta\left|\psi_{n}\right|^{2}\psi_{n}\label{eq:shrodi}\end{equation}
Where $\epsilon_{n}$ are random potentials chosen uniformly from
the interval $\left[-2,2\right]$ and $\beta$ is a positive constant.
For $\beta=0$ this is the Anderson model, where all the states are
localized. Consequently, a wave packet that is initially localized
will remain localized in the vicinity of its initial position. A question
that is subject to extensive research is whether Anderson localization
can survive the nonlinear term $\beta\left|\psi\right|^{2}\psi$ \cite{Shepelyansky1993}.
Numerical simulations indicate that for \eqref{eq:shrodi} Anderson
localization is destroyed and subdiffusion takes place \cite{Flach2009,Pikovsky2008,Molina1998,Skokos2009,FlachErratum}.
Heuristic arguments were developed in order to explain these results
\cite{Pikovsky2008,Skokos2009,Shepelyansky1993}, but the detailed
mechanism of possible spreading is not clear. Resonances between eigenstates
of the linear model, namely \eqref{eq:shrodi} with $\beta=0$ provide
a reasonable mechanism for spreading and it is the subject of the
present paper.

Double humped states $\varphi_{+},\varphi_{-}$ are two eigenstates
of the Hamiltonian \eqref{eq:shrodi} which are localized over the
same two sites that are far in real space while their energies are
very close. An example for such states appears in Fig.1. According
to Rabi's formula, in the linear case ($\beta=0$), if one places
(at time $t=0$) a wave packet on the site of one hump and the states
are exactly symmetric or antisymmetric with respect to the interchange
of the humps, one finds the packet on the other site in time $t$
with the probability

\begin{equation}
P_{12}\left(t\right)=\sin^{2}\left(\Delta E\frac{t}{2\hbar}\right)\label{eq:rabi}\end{equation}
where $\Delta E$ is the difference between the energies of the two
double humped states and the time period is $T_{Rabi}=\frac{2\pi\hbar}{\Delta E}$.
The period is preserved also for the case when the symmetry of the
hump interchange is broken, as in the case of the random potential.
This mechanism of jumping between sites has proved to be the main
mechanism for low frequency ac conductivity in disordered media \cite{Sivan1987}.
It is expected that its behavior may be strongly affected by the nonlinear
term. For a double-well potential the low energy states are symmetric
and antisymmetric double humped states, with the humps in the centers
of the wells. In the absence of nonlinearity, \eqref{eq:rabi} holds.
But, for sufficiently strong nonlinearity the wavepacket will be confined
to the initial well \cite{Soffer2005,Sacchetti2009}. In the present
work we would like to explore if double humped states contribute to
a mechanism of resonant spreading in the NLSE.

\begin{figure}
\includegraphics[width=10cm,height=7.5cm]{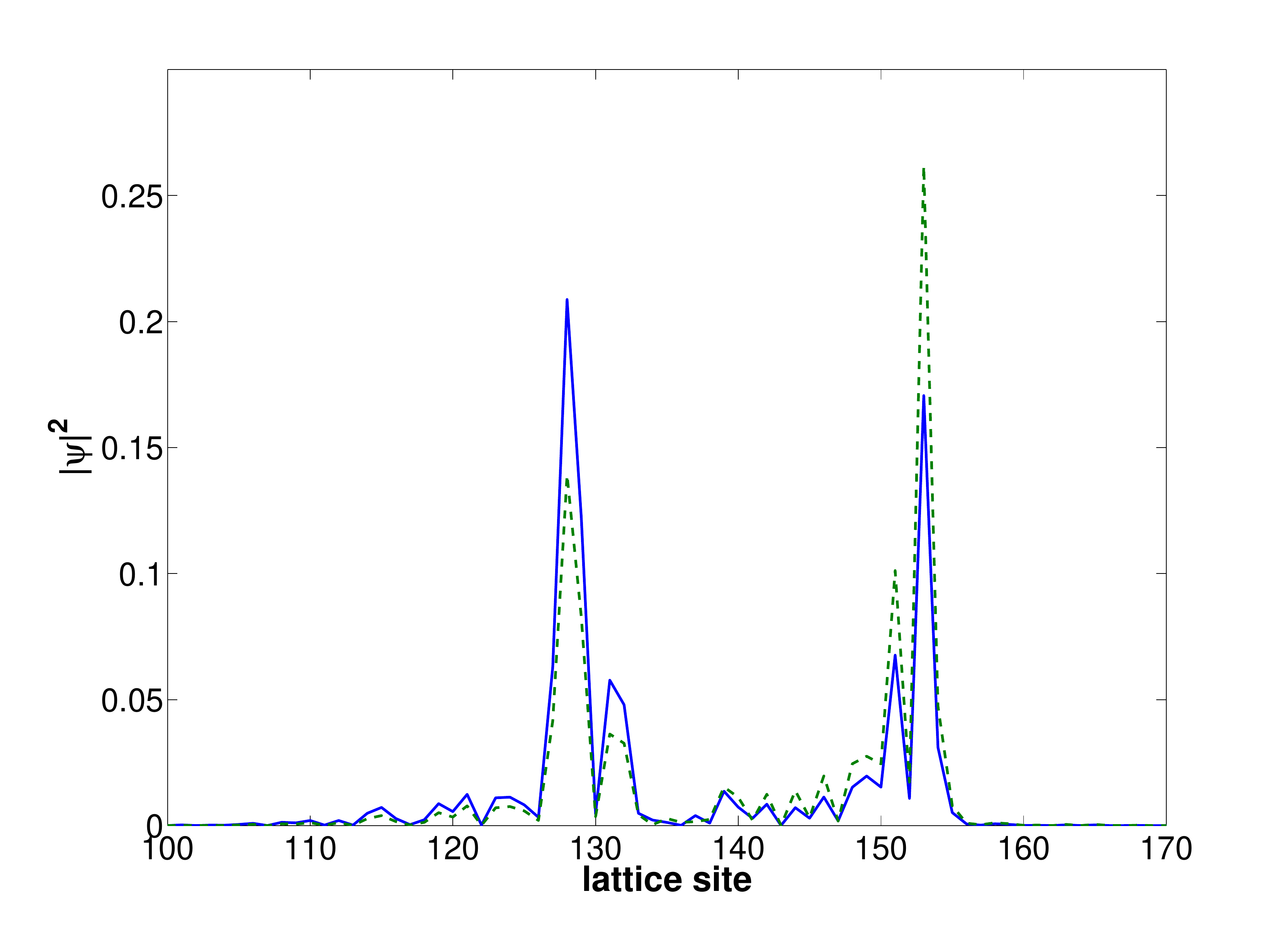}\caption{A pair of double humped states for the linear system ($\beta=0$).
The states are marked with blue solid line and green dashed line.}

\end{figure}

In our study, we would like to distinguish between resonant spreading
(caused by the double humped states) and diffusive spreading from
one state to its neighbors. For this purpose, we had to find realizations
where the humped states are located far from each other (in comparison
with the localization length, in our case $\xi\approx6$). The probability
to find double humped states with humps located at a distance $L$
is proportional to $\exp\left(-L/\xi\right)$ \cite{Sivan1987} and
therefore realizations which couple states located far from each other
are very rare (as we are looking for them in a finite region in real
space). In order to overcome this problem and create a pool of realizations
with double humped states in some region in real space, we have developed
a strategy for {}``double humps hunting''. We choose some random
potential having localized eigenstates in the linear case (Anderson
localization). We focus on two sites so that we will have double humped
states which are localized on these two sites. These sites will be
denoted by $O$ and $P$ in what follows. The Hamiltonian of this
realization is diagonalized, which results in a diagonal matrix with
eigenenergies on the diagonal. Now, we vary the site energy of the
original model on one site (say $P$) of the two sites mentioned above.
According to Feynman-Hellman theorem, when we increase monotonically
the potential of a site, its energy is monotonically increasing and
we can easily find a point where the two diagonal terms are approximately
equal. Since we change the realization, the Hamiltonian is not diagonal
anymore and the sites are coupled by matrix elements of the order
of $\exp\left(-L/\xi\right)$. Taking the potential realization which
creates almost identical energies in the diagonal of the Hamiltonian
(written in the initial eigenstates basis), we can usually construct
double humped states for this realization. In this way, we found a
set of realizations having double humped states with distance of 25
sites (about 4 localization lengths in our case) between the humped
sites.

After choosing appropriate realizations, we had to know which values
of $\beta$ should be chosen in order to see the influence of the
double humps. If we choose very small values of $\beta$, the system
will behave similarly to the linear case and a wave packet initially
localized on one humped site will oscillate between the humped sites
for very long times. However, for large values of $\beta$, the linear
eigenstates become irrelevant very quickly (compared to the period
of the oscillations) and the correlation between the double humped
states is broken before they have a chance to affect the dynamics.
Moreover, high values of $\beta$ suppress the oscillations between
the humped states even in the double-well case \cite{Sacchetti2009}
where there is no mixing with other states. So, we have to choose
the $\beta$ values very carefully. For this purpose, we use a double-well
model \cite{Tsukada2001} to set the scale of the effect of $\beta$.
In particular we find numerically for each disorder realization a
value of $\beta_{\frac{1}{4}}$ for which only $\frac{1}{4}$ of the
wave function oscillates between the humped sites $O$ and $P$ when
the double humped states are detached in the computation from all
other states. When we run the dynamics of \eqref{eq:shrodi} for double
humped realizations with $\beta=\beta_{\frac{1}{4}}$, we can see
clearly the influence of the resonance and we are still able to observe
spreading for reasonable times.

In other words, the nonlinearity has two effects: destroying the double
humped states and populating other states of the linear model. In
order to distinguish the two effects we compare to the double-well
model with a nonlinear term, where the two lowest energy states can
be assumed isolated from the other states.

The difference between the double-well model and \eqref{eq:shrodi}
is that in the double-well model only two states participate in the
dynamics (see appendix) and only these were taken into account for
this model. Therefore, numerical calculations for the double-well
model are much faster and allow us to estimate the behavior of \eqref{eq:shrodi}
without performing time consuming (split-step) calculations. In addition,
the dynamics in the double-well problem is periodic and gives us the
time scale of the oscillations. Deviations of \eqref{eq:shrodi} from
the double-well model appear when additional states become involved
in the dynamics. This happens, naturally, when we increase $\beta$.
So, first we should calculate $\beta_{c}$, the largest $\beta$ for
which the double-well model dynamics is still similar to \eqref{eq:shrodi}
and make sure that $\beta_{c}>\beta_{\frac{1}{4}}$ (otherwise, our
results for $\beta_{\frac{1}{4}}$ will have no clear meaning for
the NLSE). We have located an initial wave packet $\overrightarrow{y_{O}}$
around one of the humps (as a superposition of the two double humped
states) at site $O$. We have followed the population difference between
the double humped states in the double-well model and in the NLSE
during one time period $T$ which is numerically calculated for each
realization based on the nonlinear double-well model (see appendix).
$\beta_{c}$ was defined as the highest $\beta$ value for which $\frac{1}{T}\int_{0}^{T}\left(w_{double-well}-w_{NLSE}\right)^{2}dt<0.001$
where $w$ denotes the population difference. $\beta_{c}$ is expected
to be high when the overlap between the double humped states and other
states in the system is small and we can see a correlation between
$\beta_{c}$ and the parameter\begin{equation}
R^{-1}=\sum_{i}{}^{'}\left|\frac{V_{i}^{OOO}}{E_{O}-E_{i}}\right|\label{eq:R}\end{equation}
The index $i$ in the sum runs over all the eigenstates on the lattice
except for the two double humped states. $E_{O}$ is the energy of
the initial wave localized at site $O$ and $E_{i}$ are the eigenvalues
of the system. The numerator is $V_{i}^{OOO}\equiv\sum_{n}y_{O,n}^{3}\cdot v_{i,n}$
where the eigenfunctions with center of localization at site $i$
are denoted by $\overrightarrow{v_{i}}$ and $v_{i,n}$ is the $n$
component of vector $\overrightarrow{v_{i}}$ while $y_{O,n}$ is
the $n$ component of $\overrightarrow{y_{O}}$. The reasoning for
the importance of \eqref{eq:R} is explained in \cite{Flach2009,Skokos2009,FlachErratum}
(where $R$ is defined in a slightly different way) and the correlation
to $\beta_{c}$ is shown in Fig. 2. The correlation deteriorates when
$V_{i}^{OOO}$ is replaced by other quartic combinations of components
of $\overrightarrow{y_{O}}$ and $\overrightarrow{v_{i}}$.

\begin{figure}
\includegraphics[width=10cm,height=7.5cm]{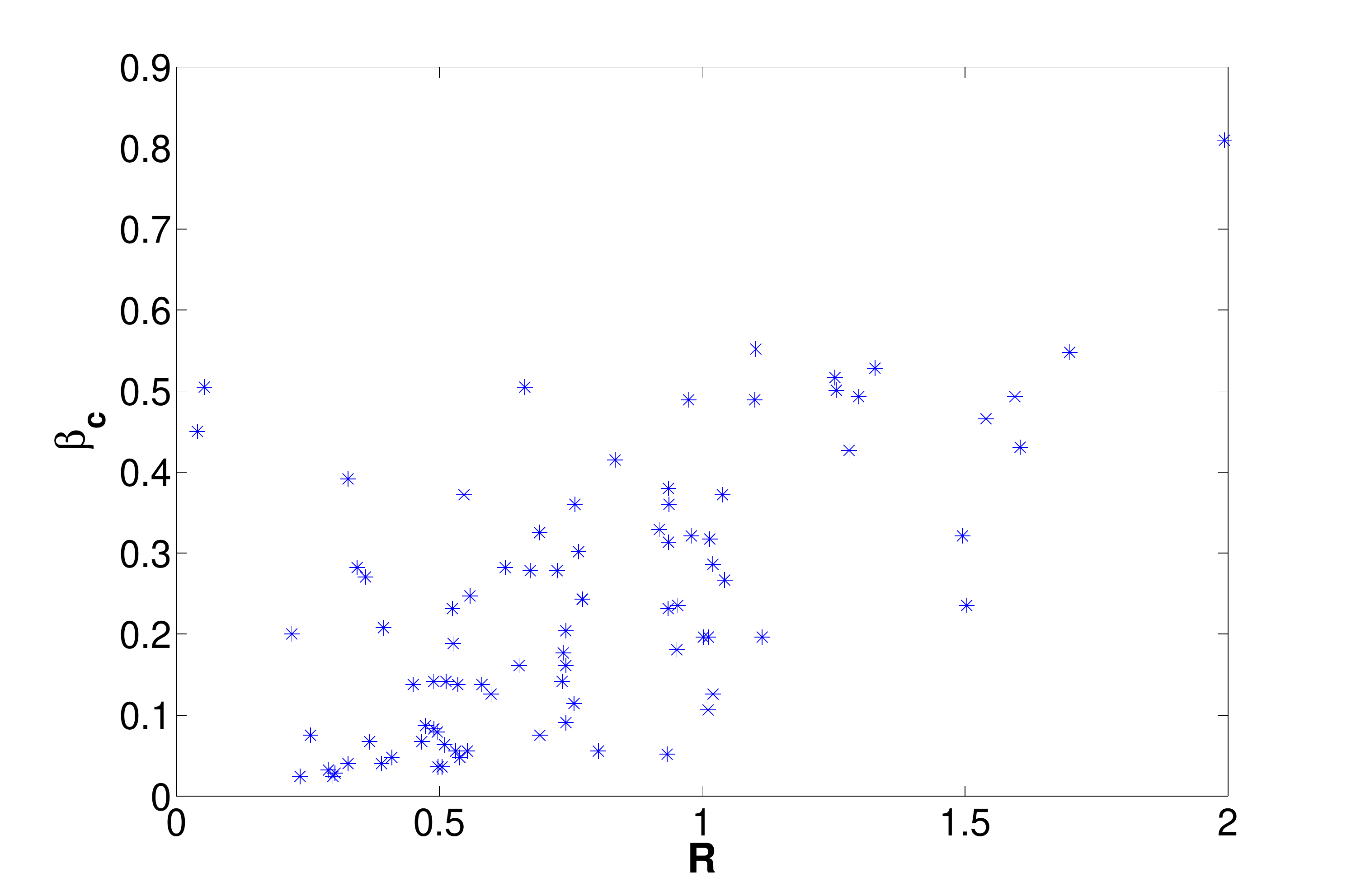}\caption{Correlation between $R$ and $\beta_{c}$ (blue asterisks).}

\end{figure}

In order to see the influence of the double humped states in specific
realizations, we should compare them to realizations where such states
are broken that will be named {}``broken realizations''. The {}``broken
realizations'' are the same set of realizations as the realizations
where double humped states are found except for one fundamental difference
- we have changed the disorder potential in one of the humped sites
to be zero, namely $\epsilon_{P}=0$, and by this broke the coupling
and destroyed the double humped states without causing qualitative
changes to the other eigenstates of the system. For each pair of double
humped and {}``broken'' realizations, we have chosen an initial
wavefunction $\overrightarrow{y_{O}}$ located around $O$ as a superposition
of the two double humped states and followed the evolution of the
wavefunction. The evolution of the wavepacket in time was calculated
according to \eqref{eq:shrodi} with $\beta=\beta_{\frac{1}{4}}$
using the split step method \cite{Skokos2009}. A quantity which interests
us when we measure the spreading of a wavefunction is the second moment,
defined as\begin{equation}
m_{2}=\sum_{n}\left(n-\bar{n}\right)^{2}\left|\psi_{n}\right|^{2}\end{equation}
where $\overline{n}=\sum_{n}n\left|\psi_{n}\right|^{2}$ is the averaged
location of the wavefunction. When we compare the growth in the second
moment for double humped realizations and the broken realizations,
we see that the second moment of the double humped realizations grows
faster, when the realizations are selected as was outlined above and
in both cases the initial wave packet is localized at $O$. Some examples
are presented in Fig. 3. We examined 25 realizations of this form
and the behavior presented in Fig. 3 is representative of all of them.
This indicates that double humped states do substantially contribute
to the spreading process of a wavefunction more than typical states.

In conclusion, we see that in the presence of nonlinearity that is
not too strong, the spreading of a wave packet prepared initially
near some site $O$ is substantially stronger if there is a double
humped state with one of its humps near $O$, than if the states peaked
near $O$ are single humped. We found that there is a regime of values
where $\beta$ is sufficiently small so that the double humped structure
is preserved but the packet is not only oscillating between the humps
but also leaks to other states, leading to spreading. In order to
find this nonlinearity regime, we have used the double-well model
to isolate the two double humped states from the other eigenstates
of \eqref{eq:shrodi} with $\beta=0$. We found that if $\beta$ is
small enough so that the oscillations between the two states are not
suppressed in the double-well model, then the double humped states
will contribute to the spreading for the NLSE. Since double humped
states are suppressed and do not contribute to the spreading for high
nonlinearities, we can not conclude that they dominate the spreading
of the NLSE. Exploring what is the dominant mechanism for this problem
is left for future a research.
\begin{acknowledgments}
Acknowledgments

We had interesting discussions with S. Aubry, S. Flach, I. Guarneri,
D. Krimer, A. Pikovsky, Ch. Skokos, D.L. Shepelyansky and A. Soffer.
This work was partly supported by the Israel Science Foundation (ISF),
by the US Israel Binational Science Foundation (BSF), by the Minerva
Center of Nonlinear Physics of Complex Systems, by the Shlomo Kaplansky
academic chair and by the Fund for promotion of research at the Technion.
The work was done partially while the authors visited the Max Planck
Institute in Dresden in March 2009, and enjoined the hospitality of
S. Flach.
\end{acknowledgments}
\begin{figure}

\includegraphics[width=15cm,height=10cm]{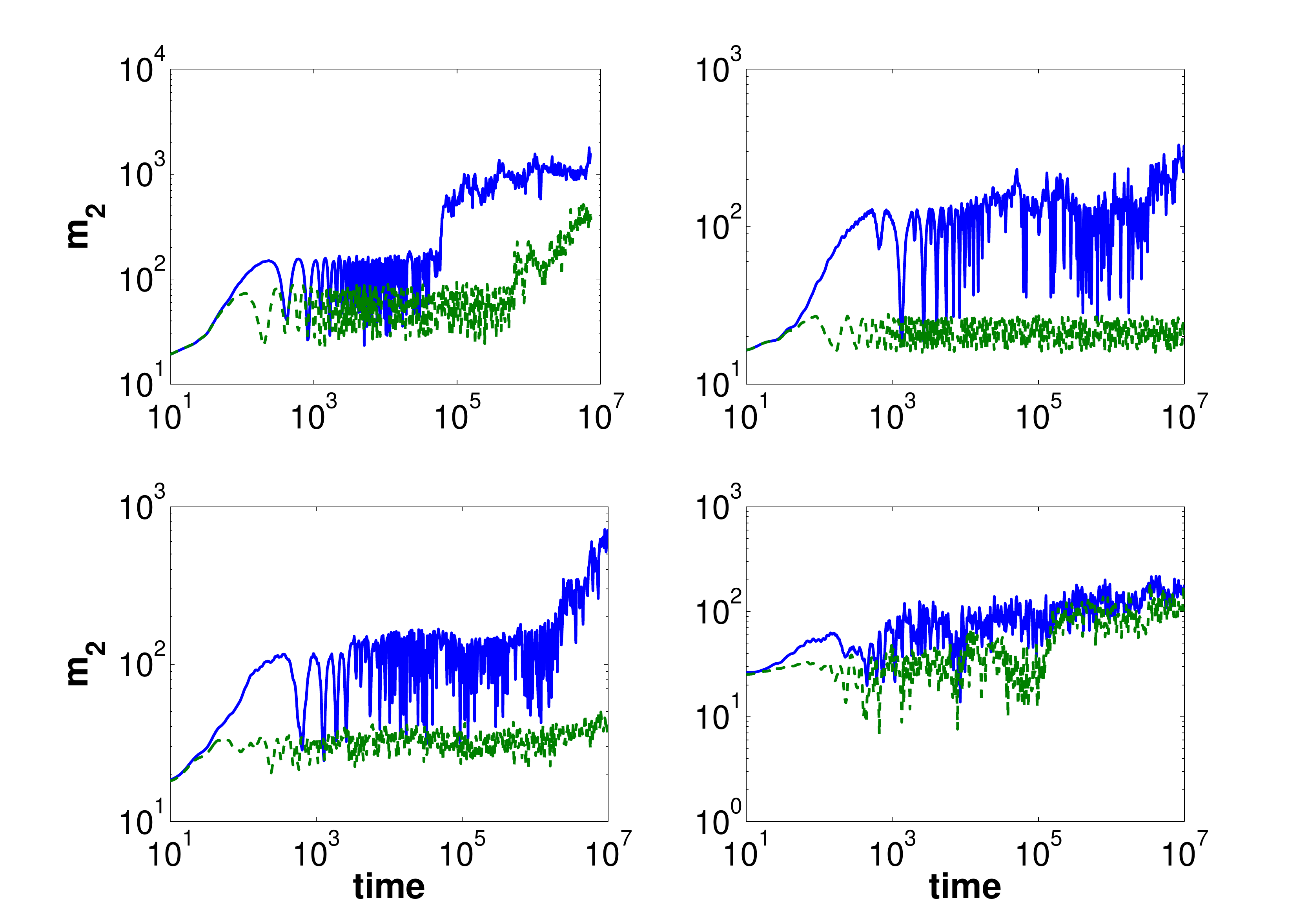}\caption{The second moment as function of time for a representative double
humped (solid blue) and broken (dashed green) realizations for wave
packets started in the vicinity of $O$.}

\end{figure}

\section*{Appendix- Double-well model}

In order to predict the response of the double humped states to variations
of $\beta$, we first investigate a model where only two states exist,
the double-well model. In this way we avoid the influence of the other
states of the NLSE. For this model, the NLSE is\begin{equation}
i\frac{\partial\Psi\left(r,t\right)}{\partial t}=-\nabla^{2}\Psi\left(r,t\right)+\left[\epsilon\left(r\right)+\beta\left|\Psi\left(r,t\right)\right|^{2}\right]\Psi\left(r,t\right)\label{eq:ap1}\end{equation}
where $\epsilon\left(r\right)$ is the double-well potential. It is
convenient to write the wavefunction in the form \cite{Raghavan1999}
\begin{equation}
\Psi\left(r,t\right)=\psi_{1}\left(t\right)\phi_{1}\left(r\right)+\psi_{2}\left(t\right)\phi_{2}\left(r\right)\end{equation}
where $\phi_{1}\left(r\right)$ and $\phi_{2}\left(r\right)$ are
symmetric and antisymmetric combinations of the double humped eigenstates
(and therefore they are orthogonal) while $\psi_{1}\left(t\right)$,
$\psi_{2}\left(t\right)$ are the amplitudes of $\phi_{1}\left(r\right)$,
$\phi_{2}\left(r\right)$ at time $t$. Eq. \eqref{eq:ap1} takes
the form:\begin{equation}
i\left[\phi_{1}\frac{d\psi_{1}}{dt}+\phi_{2}\frac{d\psi_{2}}{dt}\right]=-\left[\psi_{1}\nabla^{2}\phi_{1}+\psi_{2}\nabla^{2}\phi_{2}\right]+\left[\epsilon\left(r\right)+\beta\left|\Psi\left(r,t\right)\right|^{2}\right]\Psi\left(r,t\right)\label{eq:schrodi}\end{equation}
After multiplying both sides by $\phi_{1}\left(r\right)$ ($\phi_{1}$
and $\phi_{2}$ are localized and therefore they can be chosen to
be real) and integrating over $r$, \eqref{eq:schrodi} becomes\begin{eqnarray}
i\frac{d\psi_{1}}{dt} & = & -\int\left[\psi_{1}\phi_{1}\nabla^{2}\phi_{1}+\psi_{2}\phi_{1}\nabla^{2}\phi_{2}\right]dr+\int\epsilon\left[\phi_{1}^{2}\psi_{1}+\phi_{2}\phi_{1}\psi_{2}\right]dr\label{eq:monster}\\
 & + & \beta\int dr\left[\psi_{1}^{2}\psi_{1}^{*}\phi_{1}^{4}+\left(\psi_{1}^{2}\psi_{2}^{*}+2\left|\psi_{1}\right|^{2}\psi_{2}\right)\phi_{1}^{3}\phi_{2}+\right.\nonumber \\
 & + & \left.\left(2\left|\psi_{2}\right|^{2}\psi_{1}+\psi_{2}^{2}\psi_{1}^{*}\right)\phi_{1}^{2}\phi_{2}^{2}+\left|\psi_{2}\right|^{2}\psi_{2}\phi_{2}^{3}\phi_{1}\right]\nonumber \end{eqnarray}
Following \cite{Tsukada2001}, it is convenient to write \eqref{eq:monster}
as\begin{eqnarray}
\frac{d\psi_{1}}{dt} & = & -i\left(\omega_{1}+\Omega_{1}\left|\psi_{1}\right|^{2}\right)\psi_{1}-iK\psi_{2}\label{eq:phi1}\\
 &  & -i\left(2A_{1}\psi_{2}+A_{1}\psi_{1}^{2}\psi_{2}^{*}+B\psi_{2}^{2}\psi_{1}^{*}+A_{2}\left|\psi_{2}\right|^{2}\psi_{2}-2A_{1}\left|\psi_{2}\right|^{2}\psi_{2}\right)\nonumber \end{eqnarray}
where \begin{equation}
\omega_{1}=-\int\left(\left|\nabla\phi_{1}\right|^{2}+\epsilon\phi_{1}^{2}+2\beta\phi_{1}^{2}\phi_{2}^{2}\right)dr\label{eq:omega1}\end{equation}
 \[
\Omega_{1}=-\beta\int\left(\phi_{1}^{4}-2\phi_{1}^{2}\phi_{2}^{2}\right)dr\]
 \[
K=-\int\left(\nabla\phi_{1}\nabla\phi_{2}+\epsilon\phi_{1}\phi_{2}\right)\]
 \[
A_{1}=-\beta\int\phi_{1}^{3}\phi_{2}dr\]
 \[
A_{2}=-\beta\int\phi_{2}^{3}\phi_{1}dr\]
 \[
B=-\beta\int\phi_{1}^{2}\phi_{2}^{2}dr\]
 and we have used the relation $\left|\psi_{1}\right|^{2}+\left|\psi_{2}\right|^{2}=1$.
In a similar way,\begin{eqnarray}
\frac{d\psi_{2}}{dt} & = & -i\left(\omega_{2}+\Omega_{2}\left|\psi_{2}\right|^{2}\right)\psi_{2}-iK\psi_{1}\label{eq:phi2}\\
 &  & -i\left(2A_{2}\psi_{1}+A_{2}\psi_{2}^{2}\psi_{1}^{*}+B\psi_{1}^{2}\psi_{2}^{*}+A_{1}\left|\psi_{1}\right|^{2}\psi_{1}-2A_{2}\left|\psi_{1}\right|^{2}\psi_{1}\right)\nonumber \end{eqnarray}
where \begin{equation}
\omega_{2}=-\int\left(\left|\nabla\phi_{2}\right|^{2}+\epsilon\phi_{2}^{2}+2\beta\phi_{1}^{2}\phi_{2}^{2}\right)dr\label{eq:omega2}\end{equation}
 and \[
\Omega_{2}=-\beta\int\left(\phi_{2}^{4}-2\phi_{1}^{2}\phi_{2}^{2}\right)dr\]
In order to establish the connection with the double humped states
of \eqref{eq:shrodi}, the coefficients of \eqref{eq:omega1} and
\eqref{eq:omega2} were taken from the the Schr\"{o}dinger Eq. \eqref{eq:shrodi}.
First we have expressed these coefficients for the linear case $\beta=0$
where only $\omega_{1}$, $\omega_{2}$ and $K$ do not vanish. For
this purpose we find $\varphi_{+}$ and $\varphi_{-}$, the double
humped eigenstates of \eqref{eq:shrodi} for $\beta=0$. the amplitudes
$\psi_{1}\left(t\right)$ and $\psi_{2}\left(t\right)$ of the symmetric
and antisymmetric combinations\begin{equation}
\phi_{1}\left(r\right)=\frac{1}{\sqrt{2}}\left(\varphi_{+}+\varphi_{-}\right)\label{eq:1}\end{equation}
\begin{equation}
\phi_{2}\left(r\right)=\frac{1}{\sqrt{2}}\left(\varphi_{+}-\varphi_{-}\right)\label{eq:2}\end{equation}
satisfy the Schr\"{o}dinger Eqs. \eqref{eq:phi1} and \eqref{eq:phi2}.
Therefore, When we write the Hamiltonian \eqref{eq:shrodi} in a basis
composed from $\phi_{1}$ and $\phi_{2}$ in addition to all the single
humped eigenstates of \eqref{eq:shrodi}, $K$ will appear as an off
diagonal term which couples $\phi_{1}$ and $\phi_{2}$ while $\omega_{1}$
and $\omega_{2}$ will appear as diagonal terms. In the nonlinear
case, $K$ stays the same while $\omega_{1,2}=\omega_{1,2}^{linear}-2\beta\int\phi_{1}^{2}\phi_{2}^{2}dr$.
In order to find the corrections to $\omega_{1,2}$ for the nonlinear
Hamiltonian and to calculate all the other coefficients \eqref{eq:omega1}
and \eqref{eq:omega2}, we find numerically the vectors $\phi_{1}$
and $\phi_{2}$ with the help of \eqref{eq:1} and \eqref{eq:2} from
\eqref{eq:shrodi} as explained above.

It is convenient to follow the dynamics described by the variables
$u=\psi_{1}\psi_{2}^{*}+\psi_{2}\psi_{1}^{*}$, $v=-i\left(\psi_{1}\psi_{2}^{*}-\psi_{2}\psi_{1}^{*}\right)$
and $w=\left|\psi_{1}\right|^{2}-\left|\psi_{2}\right|^{2}$. After
some simple procedures, we obtain a vector equation of the motion
\cite{Tsukada2001},\begin{equation}
\frac{d\overrightarrow{\rho}}{dt}=\overrightarrow{\rho}\times\overrightarrow{T}\label{eq:rotation}\end{equation}
in which $\overrightarrow{\rho}=\left(u,v,w\right)$ is a vector characterizing
the state of the coupled system on the unit sphere, i.e., $u^{2}+v^{2}+w^{2}=1$,
and, $\overrightarrow{T}=\left(T_{1},T_{2},T_{3}\right)$ where\begin{equation}
T_{1}=\omega_{1}-\omega_{2}+\frac{1}{2}\Omega_{1}\left(1+w\right)-\frac{1}{2}\Omega_{2}\left(1-w\right)+\left(A_{1}-A_{2}\right)u\end{equation}
 \begin{equation}
T_{2}=-Bv\end{equation}
\begin{equation}
T_{3}=2K+2\left(A_{1}+A_{2}\right)+Bu-A_{1}\left(1-w\right)-A_{2}\left(1+w\right)\end{equation}
It is easy to find (numerically) $w\left(t\right)$ which gives us
the time period and the amplitude of the double-well oscillations.
For small nonlinearities, these results are good estimations of the
NLSE behavior on a lattice. In this work, we have used \eqref{eq:rotation}
to find $\beta_{\frac{1}{4}}$ values. For this purpose, we have chosen
an initial value for $w$ which represents a wavepacket localized
around site $O$. Following the dynamics of $w(t)$ for different
values of nonlinearity $\beta$, we have found the maximal $\beta$
for which at least $\frac{1}{4}$ of the wavepacket is oscillating
between sites $O$ and $P$. This $\beta$ is $\beta_{\frac{1}{4}}$.

\end{document}